\documentclass[a4paper]{article}
\usepackage{epsfig}

\usepackage{amsfonts}
\usepackage{bbm}

\begin{document}


\title{Partial Observers}\author{Thomas Marlow\\ \emph{School of Mathematical Sciences, University of Nottingham,}\\
\emph{UK, NG7 2RD}}

\maketitle

\begin{abstract}
We attempt to dissolve the measurement problem using a quasi-anthropic principle which allows us to invoke rational observers.  We argue that the key feature of such observers is that they are rational (we need not care whether they are `classical' or `macroscopic' for example) and thus, since quantum theory can be expressed as a rational theory of probabilistic inference, the measurement problem is not a problem.
\end{abstract}


Arguably, the basic measurement problem in quantum theory is the fact that observers don't seem to be passive when it comes to the quantum domain.  They seem to weirdly cause things to happen.  This is obviously very difficult to come to terms with (and people have admirably tried to come to terms with it).  We believe, however, that it is not something that we have to come to terms with and we shall outline in a moment why we don't.

These measurement problems seem to arise when we introduce something that is not explained by quantum theory.  For example, in the orthodox interpretation we introduce `classical' observers,  in decoherence theory we introduce `macroscopic' objects,  in other interpretations we invoke `conscious' observers that cause wavefunctions to collapse or properties to actualise, and in the many worlds interpretation we say that the `multiverse' splits into many `universes' such that our experience is explained by the mere fact that we exist in a `universe' and not in the `multiverse'.  All these attempts to explain away the measurement problem invoke a significant amount of metaphysical baggage which is part and parcel of the problem.  In the following discussion we will also introduce something that is not necessarily explained by quantum theory, but that `something' will have minimal spurious metaphysical baggage.

We will assume something analogous to the anthropic principle;  namely we claim that we exist and that we also try to be rational.  So, instead of assuming observers are `classical, `macroscopic', `conscious' or that they live in a `multiverse', let us merely presume that they, like us, try to be rational.  On the face of it, one might think that `rational' is as loaded a term as the others listed but throughout this paper we adopt the term `rational' only as it is used in common parlance---we do not consider it something uniquely human, absolute or metaphysical.  We can try to define criteria of rationality that these rational observers ought to obey.  Then, perhaps, we can derive the probabilistic side of the quantum formalism from these desiderata thus dissolving the measurement problem---there would be nothing `invasive' about rational observers' probabilistic predictions.  The probabilities that these rational observers predict will merely be assignments that they make, and they will make these assignments passively.  We balk at the idea that there is a dualist split between the quantum realm and the classical realm, or between the microscopic and macroscopic, or between our universe and the multiverse, but clearly we need not balk at the idea of some gestalts of quantum sub-systems having a certain behaviour that results in them communicating between each other in a manner that might be called `rational'.  This is close to Bohr's interpretation on the matter \cite{BohrBOOK} except that the important property of observers is not that they are somehow `classical' but that they are (or try to be) `rational'.

Clearly, such observers, in the case of quantum theory, will not be party to sufficient information to make an informed truth assignment where all relevant propositions are either `true' or `false' (this is the Kochen-Specker theorem \cite{KS}) and instead they must make informed probability assignments.  We call these observer's `partial' to emphasise this---they are party to less information than they would need to make informed truth assignments.

Partial observers can analyse one quantum system using its interaction with another quantum system \cite{Rovel96}.  We cannot justify any claim that these observers affect the world around them by `rationalising'.  These rational agents rationalise using their brains and they need not yet, when they are being rational, claim that the workings of their brains affect the world at large \emph{except through their actions}.  Thus these rational agents should reject any claim that they affect the world at large merely through being `conscious', `macroscopic', `classical' or `in only one branch of the multiverse'.  Thus when a rational observer `rationalises' and makes predictions this is \emph{exactly all that she does}.  If such observers are partial then their predictions obviously depend upon the information that they are party to. Nor do two partial observers making different predictions cause different things to happen---exactly all that they do is make different predictions, presumably because they are party to different information.

However, are we skirting around a harder measurement problem?  Clearly these partial observers \emph{will} affect the world at large when they go about checking whether their predictions are apt.  However, surely such `invasivity' is uncontroversial?  Partial observers are gestalts of quantum sub-systems and their interactions with the world at large might be analysed wholly quantum mechanically (from the perspective of a \emph{different} partial observer).  It might be that quantum theory is merely a theory of the probabilistic predictions that rational partial observers make in regards to the \emph{interactions} between quantum systems.  As long as we presume that quantum theory is no more than this then the measurement problem is dissolved.  So, there is nothing to stop partial observer $A$ investigating the probabilistic features of the interaction between another partial observer $B$ treated quantum mechanically and the quantum system that $B$ is interacting with, it is merely a very complicated thing to do.  However, all predictions will always be with respect to a particular partial observer (or a set of partial observers working intersubjectively).  In principle, $A$ can either use orthodox or Bayesian statistics to check whether the interaction between $B$ and the quantum system $B$ investigates embodies the statistical properties that $A$ predicts using quantum mechanics.

Similarly, can partial observers investigate the interactions between the quantum subsystems that make up other partial observers?  Of course we can, there is nothing to stop us investigating such a thing, it is, again, merely a very complicated thing to do.  Instead of investigating a human partial observer, one may as well start investigating something quite small and less complex.  Lets say we investigate an ant instead\footnote{This is not to say that ants rationalise---only that they are less complicated than humans.}.  A whole load of partial observers rationalising about the interactions between many small quantum subsystems of an ant might be able to come to some intersubjective agreement about the gestalt of interacting quantum systems which make up the ant.  However, there are ethical problems because it is not clear whether the set of partial observers will kill their ant in committing the team to finding out whether their predictions are apt.

The set of actions that observers can make which directly affect the world at large does not include the act of rationalising.  Clearly, observers can, and do, affect the world at large while going about and checking their predictions are apt.  As long as observers \emph{only} affect the world at large through their physical (quantum mechanical) interactions then the measurement problem is dissolved.

One response to this might be that we haven't tackled the measurement problem as it is normally expressed: in regards to there being a categorical difference between quantum dynamics and quantum measurement.  Unitary quantum dynamics predicts that a system should be in one state, and quantum measurement suggests it should be in another (an eigenstate of the measurement)---the latter supposedly not being dynamically explained by quantum theory.  This version of the measurement problem has been tackled elsewhere \cite{SR06}. It is clearly not a problem if quantum theory is regarded as a theory of partial observers' probabilistic predictions about \emph{interactions} of quantum systems.  Then states would not even be well-defined except in relation to other quantum states.  As such, the difference between a unitary state (clearly an eigenstate of \emph{some} measurement) and an eigenstate of another measurement would not be an issue; they are merely defined relative to different partial observers or quantum systems.

Another response might be that we haven't solved the problem---seemingly part of the measurement problem---that those rational observers don't see `quantum states', the world around them seems rather classical.  Nor have we tackled the problem of clearly defining how a team of partial observers might come to intersubjective agreement about the overall behaviour of a gestalt of quantum subsystems (if they each investigate the probabilistic features of a quantum interaction between a pair of the gestalt's subsystems).  These are technical problems involved in defining a classical limit of quantum theory, and they are very difficult to solve in practice.  We shall not tackle them here.

Since the key property of these observers is that they seem to be rational it is clear that we, as humans, need not be the only partial observers.  We could, in principle, work out the axioms of rationality that we use in designing a theory of probability like quantum theory and ensure that a computer also follows such axioms.  We could also, in principle, manually teach a group of humans or monkeys to do the same thing (without them necessarily being aware of it of course) by ensuring they are taught to follow simple rules which manifestly ensure that such criteria of rationality are obeyed.  Thus we would ensure that quantum theory doesn't afford us a special place in the universe---or ensure that there is necessarily a dualist split between those things that are `rational' and those that aren't.  Our first attempt at finding such criteria of rationality for quantum theory should follow an approach analogous to Cox's account of standard probability theory \cite{CoxBOOK,JaynesBOOK} (as Cox explicitly uses such criteria).  A first attempt is given in \cite{Marlow06b,Marlow06c}.

If we have a good definition of a partial observer then what are the corresponding observables that they measure?  This is a question that Rovelli compellingly answers in his wonderful `partial observables' paper \cite{Rovel02}---a paper to which we pay homage here\footnote{Although all idiosyncrasies and errors in the present paper are, of course, our own.}.

\end{document}